\documentclass{article}
\usepackage{graphicx} 
\usepackage{times,epsfig,graphics,amssymb,latexsym,amsmath,setspace}
\usepackage{array,threeparttable,hyperref,url,longtable}
\usepackage[usenames]{color}
\usepackage[authoryear,sort]{natbib}
\usepackage{paralist}
\usepackage{mathabx}
\usepackage{algorithm,algorithmic}
\usepackage{ntheorem}
\usepackage[normalem]{ulem}
\usepackage{authblk}
\usepackage[symbol*]{footmisc}
\usepackage{tabularx,ragged2e}
\usepackage{xr}
\usepackage{booktabs}

\usepackage{listings}
\usepackage[frozencache,cachedir=.]{minted}

\def\author.arg{
\medskip
Binhuan Wang \\
Data and Statistical Sciences, AbbVie Inc., Florham Park, NJ \\

\medskip
Yixin Fang \\
Data and Statistical Sciences, AbbVie Inc., North Chicago, IL \\

\medskip
Man Jin \\
Data and Statistical Sciences, AbbVie Inc., North Chicago, IL \\
}

\def\tit.arg{Statistical Inference for Chi-square Statistics or F-Statistics Based on Multiple Imputation}

\begin{document}

\pagenumbering{arabic}
\setcounter{page}{1}
\baselineskip=14pt

\begin{center}
{\Large \tit.arg} \\

\vskip 3mm

\author.arg
\end{center}

\vskip 3mm

\date{March 2024}

\begin{abstract}
Missing data is a common issue in medical, psychiatry, and social studies. In literature, Multiple Imputation (MI) was proposed to multiply impute datasets and combine analysis results from imputed datasets for statistical inference using Rubin’s rule. However, Rubin’s rule only works for combined inference on statistical tests with point and variance estimates and is not applicable to combine general F-statistics or Chi-square statistics. In this manuscript, we provide a solution to combine F-test statistics from multiply imputed datasets, when the F-statistic has an explicit fractional form (that is, both the numerator and denominator of the F-statistic are reported). Then we extend the method to combine Chi-square statistics from multiply imputed datasets. Furthermore, we develop methods for two commonly applied F-tests, Welch's ANOVA and Type-III tests of fixed effects in mixed effects models, which do not have the explicit fractional form. SAS macros are also developed to facilitate applications.

\end{abstract}

\vskip .1 in
\noindent {{\bf Key words}: \it Chi-square test; F-test; Missing data; Multiple imputation; Type-III test; Welch's ANOVA}

\doublespacing
\section{Introduction}

In medical, psychiatry, and social studies, missing data or incomplete data commonly occur due to various reasons, such as lost to follow-up and non-responses. \cite{rubin1976inference} classified missing mechanisms into three categories: missing completely at random (MCAR), missing at random (MAR), and missing not at random (MNAR). 

Multiple Imputation (MI) is one of the most popular methods for analyzing missing data. By MI, missing values in the original data set are imputed $M$ times to generate $M$ completed data sets. Then a standard statistical model is applied to each complete dataset separately. Finally, all the $M$ analytical results are combined to form a single statistical inference. \cite{rubin1986multiple} and \cite{rubin1987multiple} developed a combining rule for parameter estimates from regression analyses. SAS procedures MI and MIANALYZE are developed to implement these methods, which greatly facilitate the application of multiple imputation for researchers.

However,  Rubin’s rule only works for combined inference on statistical tests with point and variance estimates and is not applicable to combine general F-statistics or Chi-square statistics.  Then, the current SAS procedure MIANALYZE can only handle limited statistical analyses obtained using multiply imputed datasets generated by PROC MI, and hence is unable to summarize some commonly used statistical tests based on F-test or Chi-square statistics, such as Welch's ANOVA and Type-III tests of fixed effects in mixed effects models with repeated measures, when they are executed on multiply imputed datasets. 

To be specific, standard ANOVA requires the assumption of homogeneity of variance. When such an assumption is violated using statistical test such as Levene’s test \citep{levene1960robust}, Welch’s ANOVA \citep{welch1951comparison} should be used. On the other hand, in literature, there are different ways to calculate the sums of squares in order to compute F-test statistic and then p-value; as discussed in \cite{goodnight1980tests} and \cite{herr1986history}, there are at least 3 approaches, commonly called Type-I, II and III sums of squares. The Type-III test is applicable in the presence of a main effect after adjusting for the other main effects and their interactions with the main effect being tested, and therefore it should be considered when there are significant interactions. Such a Type-III test plays an important roll in testing the significance of a categorical variable with multiple categories in regression models.

In one previous paper \citep{wang2014combining}, methods and  SAS macros were developed to make PROC MIANALYZE applicable for summarizing Type-III analyses from multiple imputations, which can be applied with PROC MIXED, PROC GENMOD and PROC GLM. In this manuscript, we provide a solution to combine F-test statistics from multiple imputed datasets, when the F-statistic has an explicit fractional form (that is, both the numerator and denominator of the F-statistic are reported). Then we extend the method to combine Chi-square statistics from multiple imputed datasets. Furthermore, we develop statistical methods to combine two commonly applied F-tests based on multiple imputations, Welch's ANOVA and Type-III tests of fixed effects in mixed effects models with repeated measures, for which SAS does not provide F-test statistics with explicit fractional forms. SAS macros are developed for all above methods to facilitate applications.

The rest of the manuscript is organized as follows. Section 2 introduces the statistical methods for combining F-test statistic with an explicit fractional form and extends it to combine Chi-square statistic. In addition, statistical methods for combining two special F-test statistics without explicit fractional forms are proposed, including Welch's ANOVA and Type-III test statistic for a linear mixed effects model. Section 3 demonstrates our developed SAS macros using sample data. Section 4 finishes this paper with a brief summary. All SAS macros are provided in Appendix.

\section{Methods}

A standard multiple imputation inference involves following three distinct phases:
\begin{enumerate}
    \item The missing data are imputed $M$ times to generate $M$ complete data sets.
    \item The $M$ complete data sets are analyzed using standard procedures.
    \item The results from the $M$ complete data sets are combined for statistical inference.
\end{enumerate}
This manuscript focuses on the last step. As per Rubin's rule \citep{rubin1987multiple}, point and variance estimates for a parameter $Q$ need be computed. To be specific, let $\widehat{Q}_i$ and $\widehat{W}_i$ are point and variance estimates, respectively, from the $i$-th imputed data set, $i=1,\ldots,M$. Then we can define
$$\widebar{Q} = \frac{1}{M}\sum_{i=1}^M \widehat{Q}_i \ \textrm{and} \ \widebar{W} = \frac{1}{M}\sum_{i=1}^M \widehat{W}_i.$$
The total variance can be calculated as
$$ T = \widebar{W} + (1+\frac{1}{M})B, \ \textrm{where} \ B= \frac{1}{M-1} \sum_{i=1}^M (\widehat{Q}_i -\widebar{Q})^2.$$
Finally, the pivotal statistic for inference is $(Q-\widebar{Q})T^{-1/2}$, which is approximated distributed as $t$ with degrees of freedom $\nu_M$, where
$$\nu_M = (M-1) \left[ 1 + \frac{\widebar{W}}{(1+m^{-1})B} \right].$$

It is clear that Rubin's rule is not applicable to combine general Chi-square statistics or F-statistics.
In this section, we provide the statistical method for combining F-test statistic with an explicit fractional form  based on multiple imputation and extend it to combine Chi-square statistics. In addition, we propose methods for combining two special F-test statistics without explicit fractional forms.

\subsection{Combining F-Test Statistic with an Explicit Fractional Form} \label{sec:combine_F}

F-test is commonly utilized in conducting statistical inference, such as Type-III analyses in SAS PROC MIXED and GLM. In order to combine F-test statistics from multiple imputations, one major obstacle is how to properly combine multiple test statistics which follow F-distributions under null hypotheses. \cite{raghunathan2011analysis} established a theoretical framework of combining random variables with F-distributions in the setting of ANOVA using sum of squares.

Specifically, in an F-test statistic obtained from a complete dataset, where $s_N$ is the numerator mean squares with expectation $\sigma_N^2$ and $\nu_N$  degrees of freedom, and $s_D$ is the denominator mean squares with expectation $\sigma_D^2$ and degrees of freedom $\nu_D$. Under null hypothesis $H_0$: $\sigma_N^2 = \sigma_D^2$, we utilize the ratio
\begin{eqnarray}
\frac{P_N}{P_D} \equiv \frac{\nu_N s_N / \sigma^2_N}{\nu_D s_D / \sigma^2_D} = \frac{\nu_N s_N}{\nu_D s_D}
\end{eqnarray}
as a pivotal statistic associated with an F-distribution with degrees of freedom $(\nu_N,\nu_D)$. Based on $M$ imputed complete datasets, there are mean squares $s_N^{(l)}$ and $s_D^{(l)}$ associated with degrees of freedom $\nu_N^{(l)}$ and $\nu_D^{(l)}$, respectively, $l=1,\ldots,M$. Define 
\begin{eqnarray*} 
  A_N &=& \frac{1}{M}\sum_l \frac{1}{s_N^{(l)} }, \\
  B_N &=& \frac{1}{M}\sum_l \frac{1}{\nu_N^{(l)} s_N^{2(l)} }, \\
  C_N &=& \frac{1}{M-1}\sum_l \left( \frac{1}{s_N^{(l)}} - A_N\right)^2. 
\end{eqnarray*}
Similarly, $A_D$, $B_D$, and $C_D$ are defined for the denominator mean squares. By matching the posterior mean and variance, \cite{raghunathan2011analysis} proposed to use
\begin{eqnarray} \label{F_MI}
F_{ \textrm{\tiny MI}}= \frac{A_D}{A_N}
\end{eqnarray}
as the multiple-imputation adjusted F-statistic with the degrees of freedom $(r_N,r_D)$, where \begin{eqnarray*}
r_N &=& 2A_N^2/\left(2B_N+(M+1) C_N/M \right),\\
r_D &=& 2A_D^2/\left(2B_D+(M+1) C_D/M \right). 
\end{eqnarray*}

The above formula (\ref{F_MI}) was derived from Bayesian perspective. The basic idea is to approximate the posterior distribution of $\sigma^2_N$ and $\sigma^2_D$ by a multiple of a Chi-square distribution which matches the posterior mean and variance. Therefore, $A_N$ and $A_D$ serve as ``precision" terms which enjoy a harmonic mean form.

\subsection{Combining Chi-square statistic} \label{sec:combine_chisq}

In addition, some commonly used statistical analyses use Chi-square statistics. For example, the Type-III analyses for generalized linear models are based on the likelihood ratio test, which is associated with a Chi-square statistic rather than an F-test statistic. To be specific, $l(\beta)$ is the log-likelihood function and $l(\widehat{\beta})$ is the log-likelihood evaluated at the maximum likelihood estimate $\widehat{\beta}$. Then, maximum likelihood estimates are computed under the constraint that the Type-III function of the parameters is equal to 0, by constrained optimization. Let the resulting constrained parameter estimates be $\Tilde{\beta}$ and the log likelihood be $l(\Tilde{\beta})$. Then the likelihood ratio statistic
$$S = 2\left( l(\widehat{\beta}) -  l(\Tilde{\beta}) \right)$$
has an asymptotic Chi-square distribution under the hypothesis that the Type-III contrast is equal to 0, with degrees of freedom equal to the number of parameters associated with the effect.

Based on the statistical method proposed in Section \ref{sec:combine_F}, it is natural to extend the proposed pivotal statistic associated an F-distribution for combining F-test statistics to the case of combining Chi-square statistics. To abuse the notation, assume $\Lambda=\nu_N s_N$ is a likelihood ratio statistic associated with a Chi-square-distribution with degrees of freedom $\nu$ and note that $\sigma_N^2=1$. Therefore, we propose to use 
\begin{eqnarray}
    \chi^2_{ \textrm{\tiny MI} } = \frac{1}{A_N}
\end{eqnarray} 
as the multiple-imputation adjusted Chi-square statistic with degrees of freedom $r_N$.

\subsection{Combining F-Test Statistics without an Explicit Fractional Form} \label{sec:combine_Welch}

Sometimes, a resulting F-test statistic does not have an explicit fractional form, such as Welch's ANOVA and Type-III tests of fixed effects in mixed effects models with repeated measures. In the following two sub-sections, we proposed methods for combining results from these tests.

\subsubsection{Combining Welch's ANOVA Test} \label{sec:combine_Welch}

When the homogeneity of variances assumption required by traditional ANOVA analysis is violated, especially with unequal sample sizes, Welch's Test is a good approach for performing an ANOVA analysis. 

Define
$$    F_{ \textrm{\tiny Welch}} = \frac{\frac{1}{k-1} \sum_{j=1}^k w_j (\widebar{x}_j - \widebar{x}')^2}
    {1+\frac{2(k-2)}{k^2-1} \sum_{j=1}^k \frac{1}{n_j - 1} \left( 1-\frac{w_j}{w} \right)^2},
$$
where
$$ w_j = \frac{n_j}{s_j^2}, \ \ \ w=\sum_{j=1}^k w_j, \ \ \  \widebar{x}' = \frac{\sum_{j=1}^k w_j \widebar{x}_j}{w},$$
and then
\begin{eqnarray}
F_{\textrm{\tiny Welch}} \sim F(k-1, \gamma),
\end{eqnarray}
where
$$ \gamma = \frac{k^2}{3\sum_{j=1}^k \frac{1}{n_j - 1} \left( 1-\frac{w_j}{w} \right)^2}.$$

In order to utilize the method proposed in Section \ref{sec:combine_F}, the numerator and denominator of the $F_{ \textrm{\tiny Welch}}$ need to be calculated with associated degrees of freedom. However, SAS procedure PROC MIANALYZE dose not provide these values. In the following, we will explicitly calculate the values of numerator and denominator. Let
\begin{eqnarray*} 
  S &=& F_{\textrm{\tiny Welch}} \times (k-1) \left[ 1 + \frac{2(k-2)}{3}\gamma^{-1} \right], \\
  T &=& \gamma + \frac{2(k-2)}{3},
\end{eqnarray*}
and then 
\begin{eqnarray}
F_{\textrm{\tiny Welch}} = \frac{S/(k-1)}{T/\gamma} \sim F(k-1, \gamma).
\end{eqnarray}
Then, the method in Section \ref{sec:combine_F} can be used to conduct statistical inference based on multiple imputation.

\subsubsection{Combining Type-III Tests of Linear Combinations of Fixed Effects in Mixed Effects Models with Repeated Measures} \label{sec:combine_linearcomb}

When conducting Type-III tests of fixed effects using PROC MIXED with REPEATED or SUBJECT statement, $p$-values are computed using F-test statistics. However, these F-test statistics are based on estimable linear combinations of parameters, which cannot be written as a ratio of two quantities when the dimension of linear combination is more than one. 

One solution is to utilize Likelihood Ratio Test (LRT):
$$ \lambda_{\textrm{\tiny LR}} = 2 \ln \sup_{\theta \in \Theta} \mathcal{L}(\theta) - 2 \ln \sup_{\theta \in \Theta_0} \mathcal{L}(\theta),$$
where $\mathcal{L}(\theta)$ is the likelihood function and $\Theta_0 \subsetneq \Theta$. $\lambda_{\textrm{\tiny LR}}$ has an asymptotic Chi-square distribution with degrees of freedom equal to the difference in dimensionality of $\Theta$ and $\Theta_0$. 
For combining LRT results of Type-III tests from multiple imputation, one needs to calculate $\lambda_{\textrm{\tiny LR}}$ for each hypothesis using log-likelihood of both full model and reduced model. Then, the proposed method $\chi^2_{\textrm{\tiny MI}}$ can be applied to test each each hypothesis from multiple imputation. In SAS, it requires running PROC MIXED multiple times and data steps to manually calculate $\lambda_{\textrm{\tiny LR}}$ statistics under each hypothesis testing problem, which might be tedious and time consuming.  

One alternative method is to approximate F-distribution using other easy-to-compute distributions. \cite{li2002approximation} proposed a shrinking factor approximation (SFA) method to approximate F-distribution using Chi-square-distribution under mild conditions. Their proposed method improved the accuracy substantially that is achievable using the normal, ordinary Chi-square distribution, and Scheff\'e–Tukey approximations \citep{scheffe1944formula}. 

To be specific, let $F_{\nu_1,\nu_2}(x)$ and $G_{\nu_1}(x)$ be the cumulative distribution function (c.d.f.) of the F-distribution associated with degrees of freedom $(\nu_1,\nu_2)$ and Chi-square distribution associated with degrees of freedom $\nu_1$, respectively. Then, we can define a shrinking factor:
$$\lambda = \frac{2\nu_2 + \nu_1x/3}{2\nu_2 + 4\nu_1 x/3}.$$
For large $\nu_2$ and any fixed $\nu_1$, \cite{li2002approximation} proved
$$\sup_x | F_{\nu_1,\nu_2}(x) - G_{\nu_1}(\lambda\nu_1 x)| = O(1/\nu_2^2).$$
Numerical analysis in \cite{li2002approximation} indicates that for $n/k \geq 3$, approximation
accuracy of the SFA is to the fourth decimal place for most small values of $k$.

Therefore, to combine F-test statistic of Type-III tests associated with degrees of freedom $(\nu_1, \nu_2)$ from multiple imputation, we can use such an approximation as 
\begin{eqnarray}
    G \equiv G_{\nu_1}(\lambda \nu_1 F)
\end{eqnarray}
associated with a Chi-square distribution with a degree of freedom $\nu_1$. Then, the method in Section \ref{sec:combine_chisq} can be used to conduct statistical inference on $G$ values from multiple imputation.

\section{Example}

We demonstrate developed methods via two examples, one for Welch's ANOVA and the other for Type-III test of linear mixed effects model with repeated measures.

\subsection{Welch's ANOVA on the Sense of Smell Data}

We use the sense of smell dataset ``upsit" from the Example 50.10 in SAS online documentation for GLM procedure. There are a total of 180 subjects 20 to 89 years old in
this dataset, which are divided into five age groups. The hypothesis is that older people are more at risk for problems with their sense of smell. However, many older people also have an excellent sense of smell, which implies that the older age groups should have greater variability. Therefore, one can test the assumption of homogeneity of variance in a one-way ANOVA and conduct a Welch's ANOVA when necessary. Then, analysis using the complete dataset rejects the equal variance assumption using Levene's Test ($p$-value < 0.0001), and a Welch's ANOVA results into a $p$-value < 0.0001.

The original data set is complete, and we decide to delete the smell value of the first observation in each age group to mimic a dataset with missing values. 
First, we invoke PROC MI to generate 100 imputed datasets. Second, PROC GLM with Welch option is used to analyze each imputed dataset. Finally, the developed macro ``MIAnalyze\_Welch\_ANOVA()" is used to combine results to conduct statistical inference on the effect of age groups. The SAS scripts are shown below.

\begin{verbatim}

proc mi data=upsit_missing seed=1305417 out=out_upsit_missing 
        NIMPUTE=100;
 class agegroup;
 monotone reg (smell);
 var agegroup smell;
run;

proc glm data=out_upsit_missing;
     class agegroup;
     model smell = agegroup / ss3;
	means agegroup / welch;
     by _Imputation_;
     ods output Welch = data_Welch;
run;

%MIAnalyze_Welch_ANOVA(data_Welch);

\end{verbatim}

Table \ref{tab:example_MI_Welch} shows the final results generated by our macro. The column ``Source” shows all factors in ANOVA; ``Imputation number” shows the number of multiple imputations; ``DF” shows the degrees of
freedom of corresponding factor adjusted by the multiple imputation; ``Error DF” shows multiple imputation adjusted degrees of freedom for the error term; ``MI adjusted F”
indicates the values of F-statistics adjusted by multiple imputation; ``$p$-value” presents p-values adjusted by multiple imputation. Inference based on multiple imputation are consistent with those using the complete data.

\begin{table}[ht]
    \centering
    \begin{tabular}{cccccc}
    \toprule
      Source  & Imputation number  & DF  & Error DF  & MI adjusted F  & $p$-value  \\
      \midrule
       agegroup  & 100 & 3.97587 & 78.6885  & 14.5585 & 6.5093E-9  \\
       \bottomrule
    \end{tabular}
    \caption{Inference of Multiple Imputation Results Using Welch's ANOVA}
    \label{tab:example_MI_Welch}
\end{table}

\subsection{Type-III Test on Growth data with Repeated Measures}

We use the growth dataset ``pr" from the Example 83.2 in SAS online documentation for MIXED procedure. This dataset consists of growth measurements for 11 girls and 16 boys at ages 8, 10, 12, and 14. The analysis strategy employs a linear growth curve model for the boys and girls as well as a variance-covariance model that incorporates correlations for all of the observations arising from the same person. The working model requests an unstructured block for each SUBJECT=Person. Then, analysis using the complete dataset generates Type-III test results for Gender, Age, and Age*Gender with $p$-values of 0.2904, < 0.0001 and 0.0091, respectively.

The original data set is complete, and we decide to mimic a dataset with missing values by deleting measures at Age 14 for Person IDs 1, 5, 9, 12, 16, 20, 24, among whom first three are girls. First, we invoke PROC MI to generate 100 imputed datasets. Second, PROC MIXED is used to analyze each imputed dataset. Finally, the developed macro ``MIAnalyze\_type3\_Chisq\_approx()" is used to combine results to conduct statistical inference on the effect of Age, Gender and interaction between Age and Gender (Age*Gender). The SAS scripts are shown below.

\begin{verbatim}

proc mi data=pr_missing seed=13023587 out=out_pr_missing 
        NIMPUTE=100;
 class Gender;
 monotone reg (y = Gender Age Gender*Age);
 var Gender Age y;
run;

proc mixed data=out_pr_missing method=ml;
   class Person Gender;
   model y = Gender Age Gender*Age / s;
   repeated / type=un subject=Person r;
   by _imputation_;
   ods output Tests3 = Tests3;
run;

%MIAnalyze_type3_Chisq_approx(Tests3);

\end{verbatim}

Table \ref{tab:example_MI_type3} shows the final results generated by our macro. The column ``Source” shows all factors in Type-III tests; ``Imputation number” shows the number of multiple imputations; ``DF” shows the degrees of
freedom of corresponding factor adjusted by the multiple imputation;  ``Chisq”
indicates the values of F-statistics adjusted by multiple imputation; ``$p$-value” presents p-values adjusted by multiple imputation. $p$-values based on multiple imputation are consistent with those using the complete data.

\begin{table}[ht]
    \centering
    \begin{tabular}{ccccc}
    \toprule
      Source  & Imputation number  & DF  & Chisq  & $p$-value  \\
      \midrule
       Age  & 100 & 0.96066 & 43.2080  & 0.00000  \\
       Age*Gender & 100 & 0.21390 & 0.3759  & 0.13363  \\
       Gender & 100 & 0.00774 & 0.0000  & 0.04634  \\
       \bottomrule
    \end{tabular}
    \caption{Inference of Multiple Imputation Results Using Type-III Test with Repeated Measures.}
    \label{tab:example_MI_type3}
\end{table}

\section{Summary}

SAS procedures MI and MIANALYZE are useful and easy-to-implement tool to conduct missing data analysis via multiple imputation. However, Rubin's rule can only handle limited statistical analyses obtained using multiple imputed datasets. In this manuscript, we propose statistical methods to combine any F-test statistic with an explicit fractional form from multiple imputations and extend it to combine any Chi-square statistic. Furthermore, we develop methods for two commonly-applied F-tests, Welch’s ANOVA and Type-III test from mixed effects models with repeated measures, which do not have an explicit fractional form. All macros are also presented in Appendix to facilitate applications.

\bibliographystyle{apalike}

\bibliography{refs_type3}

\appendix

\section{Appendix}

\subsection{SAS Macro for Combining F-Test Statistic with an Explicit Fractional Form}

\begin{verbatim}


* macro for combining F-tests*;
/*
 Imputation: imputation number
 Source: factor or characters
 DF: df for factor
 SS: sum of squares
 MS: mean squares
 Error: error term or denominator
 de_DF: df of error
 MSE: mean squares of error
	
*/
%macro MIAnalyze_F_test(data_Imp_Ftest);

data &data_Imp_Ftest;
set &data_Imp_Ftest;
An=1/MS;
Bn=1/(MS**2 * DF);

Ad=1/MSE;
Bd=1/(MSE**2 * de_DF);
run;

proc means data=&data_Imp_Ftest noprint;
class source;
output out=MIanalyze mean(An Bn Ad Bd)=ave_An ave_Bn ave_Ad ave_Bd 
           var(An Ad)=ave_Cn ave_Cd max(_Imputation_)=M;
run;

data MIanalyze;
set MIanalyze;
rn=2* ave_An**2 /( 2*ave_Bn+(M+1)*ave_Cn/M );
rd=2* ave_Ad**2 /( 2*ave_Bd+(M+1)*ave_Cd/M );
MI_F=ave_Ad/ave_An;
p_value=1-PROBF(MI_F,rn,rd);
run; 

data finaloutput;
set MIanalyze;
where Source~=' ' and p_value~=.;
keep Source M rd rn MI_F p_value;
label M='Imputation number' rn=DF rd='Error DF' MI_F='MI adjusted F' 
     p_value='p-value';
run;

proc print data= finaloutput label;
where p_value~=.;
var source M rn rd MI_F p_value;
run;

%mend;

\end{verbatim}

\subsection{SAS Macro for Combining Chi-square statistic}

\begin{verbatim}


* macro for combining Chisq-tests*;
/*
 Imputation: imputation number
 Source: factor or numerator
 DF: df for factor
 Chisq: Chi-squared statistic
	
*/
%macro MIAnalyze_Chisq_test(data_Imp_Chisqtest);

data data_Imp_Chisqtest;
set &data_Imp_Chisqtest;
A=(1/ChiSq) * DF; 
B=1/(ChiSq**2) * DF;
run;

proc means data=data_Imp_Chisqtest noprint;
class source;
output out=MIanalyze mean(A B)=ave_A ave_B var(A)=ave_C max(_Imputation_)=M;
run;

data MIanalyze;
set MIanalyze;
r=2* ave_A**2 /( 2*ave_B+(M+1)*ave_C/M );
Chisq=(1/ave_A) * r;
p_value=1-PROBCHI(Chisq,r);
run; 

data finaloutput;
set MIanalyze;
where Source~=' ';
keep Source M r Chisq p_value;
label M=Imputation_number r=DF;
run;

proc print data= finaloutput label;
run;

%mend;

\end{verbatim}

\subsection{SAS Macro for Combining Welch's ANOVA Test from PROC GLM}

\begin{verbatim}


/*****************************************************
macro for combining Welch's ANOVA test
Input: ods Table Welch
Require running macro %MIAnalyze_F_test() first	
*****************************************************/

%macro MIAnalyze_Welch_ANOVA(data_Welch);

*prepare data for Welch test*;
 data data_Welch_Imp_Ftest data_Welch_error;
 set &data_Welch (drop=ProbF);
 if source='Error' then output data_Welch_error;
 else output data_Welch_Imp_Ftest;
 run;

 *rename and remove columns in the error data*;
data data_Welch_error;
set data_Welch_error;
drop Effect Dependent Source FValue;
rename DF=de_DF;/* df of the denominator */
run;

 data data_Welch_Imp_Ftest;
 merge data_Welch_Imp_Ftest data_Welch_error;
 by _Imputation_;
 k = DF + 1; /* number of groups */
 MSE = 1 + 2*(k-2)/(3 * de_DF); /* Denominator of the F-test */
 SSE = MSE * de_DF; /* Sum of squares of the denominator */
 MS = FValue * MSE; /* Numerator of the F-test */
 SS = MS * DF; /* Sum of squares of the numerator */

 drop FValue;
run;

%MIAnalyze_F_test(data_Welch_Imp_Ftest);

%mend;


\end{verbatim}

\subsection{SAS Macro for Combining Type-III Tests of Linear Combinations of Fixed Effects from PROC MIXED or GLIMMIX}

\begin{verbatim}

/****************************************************
Macro for Combining Type-III test with
Repeated or Subject statement
Input: ods table Tests3
Require running macro %MIAnalyze_Chisq_test() first
*****************************************************/

%macro MIAnalyze_type3_Chisq_approx(Tests3);


data Tests3;
set &Tests3;

*drop ProbF;

lambda = (2* DenDF + NumDF * FValue/3 + NumDF - 2)/
         (2* DenDF + 4 * NumDF * FValue/3);
ChisqValue = lambda * NumDF * FValue;
DF = NumDF;

ProbChisq = 1-PROBCHI(ChisqValue,DF);

diff_p = abs(ProbChisq - ProbF);

run;


data Tests3_sub;
set Tests3;
keep _Imputation_  Effect ChisqValue DF;

rename Effect = Source ChisqValue = Chisq;

run;

%MIAnalyze_Chisq_test(Tests3_sub);

%mend;
\end{verbatim}

\end{document}